# Collective order reorganizes dissipation and powers an active engine

Michael Riedl[1,2]*, Vincent Wattiez[3], Étienne Fodor[4], Jan Brugués[1,2]* and Francesco Romanò[3]*.

## Affiliations

[1]Cluster of Excellence Physics of Life, TU Dresden, Dresden, Germany.

[2]Max Planck Institute of Molecular Cell Biology and Genetics, Dresden, Germany.

[3]Univ. Lille, CNRS, ONERA, Arts et Metiers Institute of Technology, Lille, France.

[4]Department of Physics and Materials Science, University of Luxembourg, Luxembourg.

*Corresponding author. Email: michael.riedl@tu-dresden.de, jan.brugues@tu-dresden.de, francesco.romano@ensam.eu

## Abstract

Motile active matter systems, from animal groups to synthetic particles, exhibit spontaneous self-organized transitions between disordered and ordered collective states. These transitions are driven by continuous energy injection, yet how energy consumption and dissipation change during these transitions remains poorly understood due to the challenge of measuring individual energetic fluxes. Here, we introduce a minimal experimental system of synthetic motile spheres that enables the first time-resolved power-budget characterization of collective motion. We show that the transition to an ordered state is accompanied by increased locomotion efficiency and a reorganization of dissipation pathways, shifting dissipation from internal friction to external slip with the environment. Leveraging this energetic reorganization, we construct an active engine powered by collective motion, capable of performing mechanical work against an external load. These findings establish energetic fluxes as quantitative observables for understanding collective state transitions and demonstrate how collective order can be harnessed for work extraction in active matter.

## Main

Assemblies of interacting motile agents can spontaneously transition between distinct dynamical regimes [1–3]. A prominent example of these transitions is from disordered motion to ordered collective motion [4–6]. In confined active systems, this transition occurs often in the form of a persistent global rotation, in which the initially disordered motion gives rise to a self-organized vortex state [7–11]. Such vortex states have been observed across a wide range of systems, from driven colloids [12,13], to bacterial and cellular assemblies [14–17], to animal groups including insect swarms [18], fish schools [19], robotic systems [16,20], and even human crowds [21]. The recurrence of this self-organized state across both living and synthetic systems suggests generic underlying physical mechanisms that still remain poorly understood [22].

What links these otherwise diverse systems is a common physical feature: the individual agents sustain their motion through continuous energy injection [23]. This injected energy is redistributed among the available degrees of freedom and ultimately dissipated to the environment [24]. However, direct measurements of energy input or dissipation (e.g., metabolic rates) are experimentally challenging, particularly at the level of individuals [25]. Thus, it remains experimentally unresolved how energy input and dissipation reorganize as collectives transition between disordered and ordered states. Consequently, existing characterizations of collective transitions have primarily relied on kinematic or structural order parameters, such as velocity alignment or spatial correlations. While such descriptors successfully distinguish disordered from ordered motion [5–7], they do not detail how much energy is injected, how it is redistributed, or where it is dissipated. As a consequence, hypotheses linking coordinated motion to reduced energetic demands typically rely on indirect proxies rather than direct measurements and a description based on agent-level energetics remains unexplored [26–28].

Here, we introduce a minimal active system that allows direct measurements of energy input together with a time-resolved decomposition of dissipation, enabling an energetic characterization of collective state transitions. We combine time-resolved measurements of injected power with independent tracking of translational and rotational motion to establish a power-budget description of collective states. We show that the ordered collective state operates at a reduced energetic cost compared to the disordered state. Crucially, the distinction between order and disorder does not arise solely from a reduction in total energy consumption, but more prominently from a reorganization of dissipation pathways. These state-specific energetic signatures allow collective order and disorder to be distinguished from energetic observables alone. Moreover, the reorganization of dissipation in the ordered state allows part of the injected energy to be converted into mechanical work, enabling an active engine driven by collective motion.

## Collective order reduces power consumption at increased speed

To investigate the connection between coordinated motion and energy dissipation, we used a minimal system of active spheres that enabled direct, time-resolved measurements of power input and allowed us to resolve dissipation at the level of individual agents, while still exhibiting robust collective state transitions. Each sphere consists of a hollow shell hosting an unbalanced motor that applies a torque to an internal shaft connecting the two polar axes [16,29,30]. When placed on a rough substrate, the applied torque drives the sphere to roll along chaotic trajectories [16,30] **(Supplementary Video 1)**. To sustain the spheres' motion, each motor draws electrical power ($P_{el}$) and converts it into mechanical work ($P_{input}$). When changes in mechanical load occur the motor performs a coupled adjustment of its rotation speed and torque which directly reflects in changes in electrical power. To resolve these energetic fluxes, each sphere is equipped with a microcontroller on a compact printed-circuit board that records the instantaneous electrical power ($P_{el}$) drawn by the motor in real time and wirelessly broadcasts it. To extract the dissipation per sphere, we exploit the fact that in our system the injected electrical power per sphere equals the sum of the rates of dissipation and kinetic energies ($P_{diss}$ and $P_{kin}$, respectively). We independently resolve the translational and rotational motion of the sphere by tracking fiducial markers on the shell **(Supplementary Video 2)**. The resulting trajectories allow us to compute the corresponding time-resolved kinetic energy and thereby infer the dissipation rate per sphere. Together, these measurements characterize individual agents in terms of energy input, dissipation and changes in kinetic energies, enabling a complete power-budget analysis of their collective dynamics **(Supplementary Fig. 1, Methods)**.

We confined active spheres within a circular boundary, where populations can transition from disordered motion to a self-organized vortex state characterized by near-circular orbits and global rotation **(Fig. 1A-C) (Supplementary Video 3)**. This vortex state was previously reported for larger populations[16] and reproduced here in simulations **(Supplementary Video 4)**, and also found in other systems [14,15,17]. The transition to this state is readily visualized using kymographs of the azimuthal coordinate, revealing a change from mixed slopes in the disordered state to coherent bands with a common angular velocity in the ordered state **(Fig. 1D)**. We measured the electrical input power and translational speed in both ordered and disordered states, which we classified using a rotational order parameter **(Fig. 1E)**. The transition from disordered to collective ordered motion is accompanied by a simultaneous increase in the translational speed and a decrease in the population mean input power **(Fig. 1E - F)**. As a result, the collectively ordered state achieves higher velocities at lower energetic cost, suggesting a more efficient locomotion **(Fig. 1G)**.

**Collective order reorganizes dissipation pathways**

Our data show that the collectively ordered state injects and hence dissipates less power than the disordered state. To resolve the sources of dissipation, we decompose the total mechanical dissipation into internal dissipation ($P_{internal}$), and external dissipation due to slip at the substrate ($P_{slip}$) **(Fig. 2A)**. To quantify $P_{slip}$, we compare the rotational velocity of the shell with the translational velocity of the center of mass, the difference between these velocities directly measures the slip velocity at the substrate **(Methods)**. The total slip dissipation increases linearly with population size **(Fig. 2B)**. Closing the power balance yields the internal dissipation $P_{internal}$ from the measurements of $P_{input}$, $P_{kin}$, and $P_{slip}$. In both disordered and ordered states, the rate of change of kinetic energy $P_{kin}$ fluctuates around zero, consistent with statistically converged steady-state motion **(Fig. 2A, Supplementary Fig. 4A)**. We find that the transition from disordered to ordered states is accompanied by an increase in dissipation through slip and a concomitant decrease in internal motor dissipation **(Fig. 2A, C, Supplementary Fig. 4B - C)**. This reorganization of the dissipation is reproduced quantitatively performing fully resolved simulations of our spheres **(Supplementary Fig. 5A, Fig. 2A, B)**. Thus, while both the input power and the total dissipation decrease in the collective ordered state, the defining change is a redistribution of dissipation pathways **(Fig. 2C)**. This reorganization implies that the collective state of the system can be identified from energetic observables alone [31].

## Collective interactions suppress speed fluctuations and stabilize faster motion

Having characterized the dissipation pathways in our system, we wondered what role collective interactions play in the reorganization of dissipation we observed between the ordered and disordered states. To this end, we compared the dissipation of a population of 10 active spheres to the equivalent dissipation of non-interacting active spheres. We measured the energetics of each sphere in an identical set, first in isolation and subsequently in the disordered and ordered collective states **(Supplementary Video 5)**. Remarkably, the ordered collective state and the ensemble of independent spheres exhibited nearly identical distributions of input power **(Fig. 3A)** and dissipation modes **(Supplementary Fig. 4B - C)**, indicating that interactions in the ordered collective state do not substantially alter the instantaneous energetics. Consistent with this, simulations show that collisions between spheres act primarily as phase-shifting events that synchronize the internal dynamics of the agents, transferring only comparably small amounts of energy in the process **(Supplementary Fig. 6A - B)**.

However, we observed an unexpected increase in translational velocity for the collective that was statistically significant **(Fig. 3B)**. Since the energetic pathways are comparable in both cases, this speed enhancement cannot be attributed to changes in input energy or dissipation channels and instead could arise from differences in the temporal dynamics of motion. Supporting this interpretation, independently moving spheres display significantly larger velocity fluctuations than the collective in the ordered state **(Supplementary Fig. 6C)**. Simulations also reproduce the speed increase in the collective and reveal that it results from the collective settling into one of two ring-states at the boundary with distinct speeds; a higher-speed state and a lower-speed state **(Supplementary Fig. 6D)**. Single spheres explore a broader range of motions, leading to the experimentally observed increased velocity fluctuations. Independently moving single spheres reside less stably in the higher-speed state, which results in the overall lower mean speed **(Fig. 3B, Supplementary Fig. 6D)**. The spheres autonomously change their inclination minimizing slip at the wall and substrate, and while inter-sphere collisions in the collective stabilize the faster state, the increased speed fluctuations of isolated spheres act as a perturbation. The recurring switching between both organized states in the collective leads to pulsatile dynamics across the population and an oscillating but coordinated mechanical forcing on the substrate [16,32]. Thus, the collective ordered state behaves energetically like a set of independent agents, but dynamically as a synchronized entity with enhanced speed and suppressed speed fluctuations.

## Ordered collectives enhance mechanically transferable power

When a collective transmits forces to the environment in a coordinated and constructive manner, energy that would otherwise be dissipated can become available for mechanical work on the environment ($P_{env}$) that scales with the number of contributing agents. In our system $P_{env}$ is dominated by power transfer to the substrate ($P_{sub}$) with a smaller contribution from power transfer to the confining wall ($P_{wall}$) **(Supplementary Fig. 7A - C)**. Both contributors scale with speed, and since the collective moves faster on average than an isolated sphere, this potentially transferable power increases super-linearly with sphere number with respect to an isolated sphere **(Fig. 3C - D)**. Here, the main contributions stem from $P_{sub}$ which peaks when the outer ring is fully occupied **(Fig. 3D, Supplementary Fig. 7A)**. Together, this super-linear scaling and the synchronized, oscillatory forcing generated by cyclic acceleration and deceleration during state switching in the ordered collective raise the question of whether the collective can perform effective mechanical work and act as a driver for an active motor.

**Collective order powers an active engine**

To this end, we constructed a mechanical ratchet that rectifies the oscillating displacement into unidirectional rotation of the confinement **(Fig. 4A, Supplementary Fig. 9A)**. Placed underneath the confinement, the device consists of an axial ball bearing that allows azimuthal motion of the confinement in response to forces transmitted through the substrate. A pawl mechanism enforces the desired directional rectification, preventing motion in one direction. In this way, the device converts the collective oscillating forcing into a directed rotation of the confinement. The threshold of the required forcing for the ratchet engagement is tunable through the tooth geometry and bending stiffness of the pawl. When collective forcing exceeds this threshold, the confinement can advance in discrete steps and undergo sustained unidirectional rotation **(Supplementary Video 6)**. We quantify this rotation by tracking additional markers on the confinement **(Fig. 4B - C)**. Net ratchet displacement was observed only in the ordered state **(Fig. 4B - C)**. In this configuration, part of the mechanical energy is redirected from dissipative pathways into rotation of the confinement. Accordingly, slip-related dissipation decreases when the ratchet is engaged, indicating that a fraction of the energy injected by the collective is converted into mechanical motion of the device **(Supplementary Fig. 9B - D)**.

Adapting our power budget to the ratchet setup, we decomposed the mechanical power transmitted from the active spheres into contributions acting on the substrate and at the confining wall **(Supplementary Fig. 9B)**. The power transferred to the substrate scales with the relative velocity between the spheres and the rotating confinement. At the circular wall, centrifugal forcing transfers mechanical power that scales with the cube of the relative angular velocity **(Methods)**. Independently, we compute the mechanical power associated with the observed motion of the confinement from its angular velocity and the measured moment of inertia **(Supplementary Fig. 9, Supplementary Fig. 10) (Methods)**. The resulting mechanical power exhibits oscillations in time, reflecting pulsatile forcing by the collective **(Supplementary Fig. 9B)**. Peaks in transmitted power coincide in time with peaks in rotational power of the confinement **(Supplementary Fig. 9C)**. A stronger collective synchronization correlates with larger power spikes in confinement rotation **(Supplementary Fig. 9D)**, indicating that synchronized collective dynamics enhance mechanical power transfer. These results show that coordinated collective forcing can be rectified into directed mechanical motion.

While a motor generates motion [33–35], an engine is distinguished by its ability to perform work against an external load [36,37]. To test whether our active motor could operate as an active engine, we coupled the rotating confinement to a pulley system, with the goal of converting azimuthal motion into linear displacement of a suspended mass **(Fig. 5A-B)**. In this configuration, the ordered collective lifts the load and performs measurable mechanical work against gravity **(Fig. 5C)**. By varying the applied load, we characterized the operating range of the active engine. As expected, the displacement per cycle decreases with increasing load **(Fig. 5C)**, while the extracted mechanical power remains approximately constant over the operating regime **(Fig. 5C) (Supplementary Video 7)**. The motor stalls at a load of 10 g, defining the maximum sustainable force output under our experimental conditions. From measurements of electrical input power and mechanical output power, we determine a peak efficiency of 0.04% against a load of 7.5 g **(Fig. 5D)**. When referenced to the mechanically transferred environmental power ($P_{env}$), the effective efficiency increases to approximately 0.09%. For comparison, the efficiency of the internal electrical motor is approximately 42%, indicating that a substantial fraction of the electrical input power is already dissipated internally before mechanical energy is transmitted to the shell. Together, these results demonstrate the realization of an active engine powered by collective dynamics, in which coordinated motion is directly converted into mechanical work against an external load while preserving the ordered collective state that produces it.

## Discussion

In this study, we introduce a system that enables time-resolved characterization of the power budget of a moving collective during its self-organized transition from disordered to ordered motion. We find that the ordered state reaches higher speeds while operating at reduced energetic cost. This enhanced efficiency arises from the reorganization of the dissipation pathways, where energy dissipation switches from internal mechanical friction to external slip with the environment. This energetic reorganization provides a distinct energetic signature of the transition between ordered and disordered collective states. Finally, by building an active engine driven by the moving collective, we demonstrate that this redirection of dissipation can be exploited to perform mechanical work.

Our work establishes a complete power budget for a motile active system. Previous studies of collective behavior have primarily relied on kinematic and structural order parameters to distinguish between states [4,5]. At the same time, hypotheses suggesting that animal groups coordinate their movement to save energy have rested on indirect evidence [26,38]. Our work overcomes this limitation in a synthetic system, enabling a direct and quantitative link between energy input, dissipation, and the emergence of collective order. The finding that the ordered state operates at a reduced power input while achieving higher speeds, compared to disordered or non-interacting agents, provides a direct experimental validation that coordinated motion can be energetically advantageous.

Our work reveals that the transition from disorder to order is accompanied by an energetic signature consisting of the redistribution of dissipation from internal to external pathways [39]. This signature suggests that collective states may be distinguishable from energetic measurements alone. More generally, when a dissipative system like ours reaches an effective equilibrium in which energy input is balanced by dissipation, only states that can sustain this balance are physically accessible. This suggests that, in principle, the set of realizable collective states can be constrained a priori from the available dissipation pathways and their capacity to dispose of power. This energetic perspective on collective motion complements traditional kinematic descriptions with an energetic characterization of collective states. From this perspective, states that shift dissipation from internal losses to external interactions are particularly interesting, because external dissipation need not be wasted but can still be harnessed.

A key demonstration of the functional role of the reorganization of dissipation pathways in the collective ordered state is the realization of an active engine. By rectifying the coordinated forcing of the ordered state with a mechanical ratchet we converted otherwise dissipated energy into mechanical work. The concept of extracting work from active systems has been theorized, with proposals often focusing on rectifying fluctuations of single agents [36,37]. In contrast, our active engine relies on an emergent self-organized collective state rather than fluctuations of individual agents, providing an experimental realization of the possibility of engines powered by active matter self-organization [40].

The measured efficiency of our engine is modest, as expected for a proof-of-principle realization, but we envision improvements by optimizing motor operating points, interactions, and confinement geometry. Remote control of individual agents could further enable deliberate strategies to enhance performance [41]. Our system thus offers new avenues for designing and controlling active systems and for quantitatively testing long-standing questions about the energetic principles that govern collective biological and synthetic systems.

**Acknowledgments:** We thank Peter Traunmüller for his assistance with the design and assembly of the onboard electronics.

**Funding:** MR acknowledges funding from the Human Frontier of Science (postdoctoral cross-disciplinary fellowship LT0049/2024-L1) and the European Molecular Biology Organization (postdoctoral fellowship EMBO ALTF 227-2024). JB acknowledges funding from the European Research Council (ERC consolidator grant SynthNuc, 101045468). MR and JB acknowledge support from the Deutsche Forschungsgemeinschaft (DFG, German Research Foundation) under Germany's Excellence Strategy–EXC–2068–390729961–Cluster of Excellence Physics of Life of TU Dresden. This project has received funding from the Luxembourg National Research Fund (FNR) grant references 14389168 and 18118949. É.F. acknowledge support from Grant No. NSF PHY-2309135 to the Kavli Institute for Theoretical Physics (KITP).'

## Author contributions:

Conceptualization: MR, EF, JB, FR

Methodology: MR, JB, FR

Software: MR, VW, FR

Validation: MR, FR, JB

Formal analysis: MR, JB, VW, FR

Investigation: MR, JB, FR

Resources: MR, JB, FR

Data curation: MR, JB, FR

Writing – original draft: MR, JB

Writing – review & editing: MR, EF, JB, FR

Visualization: MR, JB, FR

Supervision: MR, EF, JB, FR

Project administration: MR, JB, FR

Funding acquisition: MR, JB, FR

**Competing interests:** Authors declare that they have no competing interests.

**Data and materials availability:** All data and code required to reproduce the analyses are available from the corresponding authors upon reasonable request and will be deposited in a public repository before publication. All other data are available in the main text or the supplementary materials.

## Supplementary Materials

Figs. S1 to S10

Supplementary Video 1 to 7

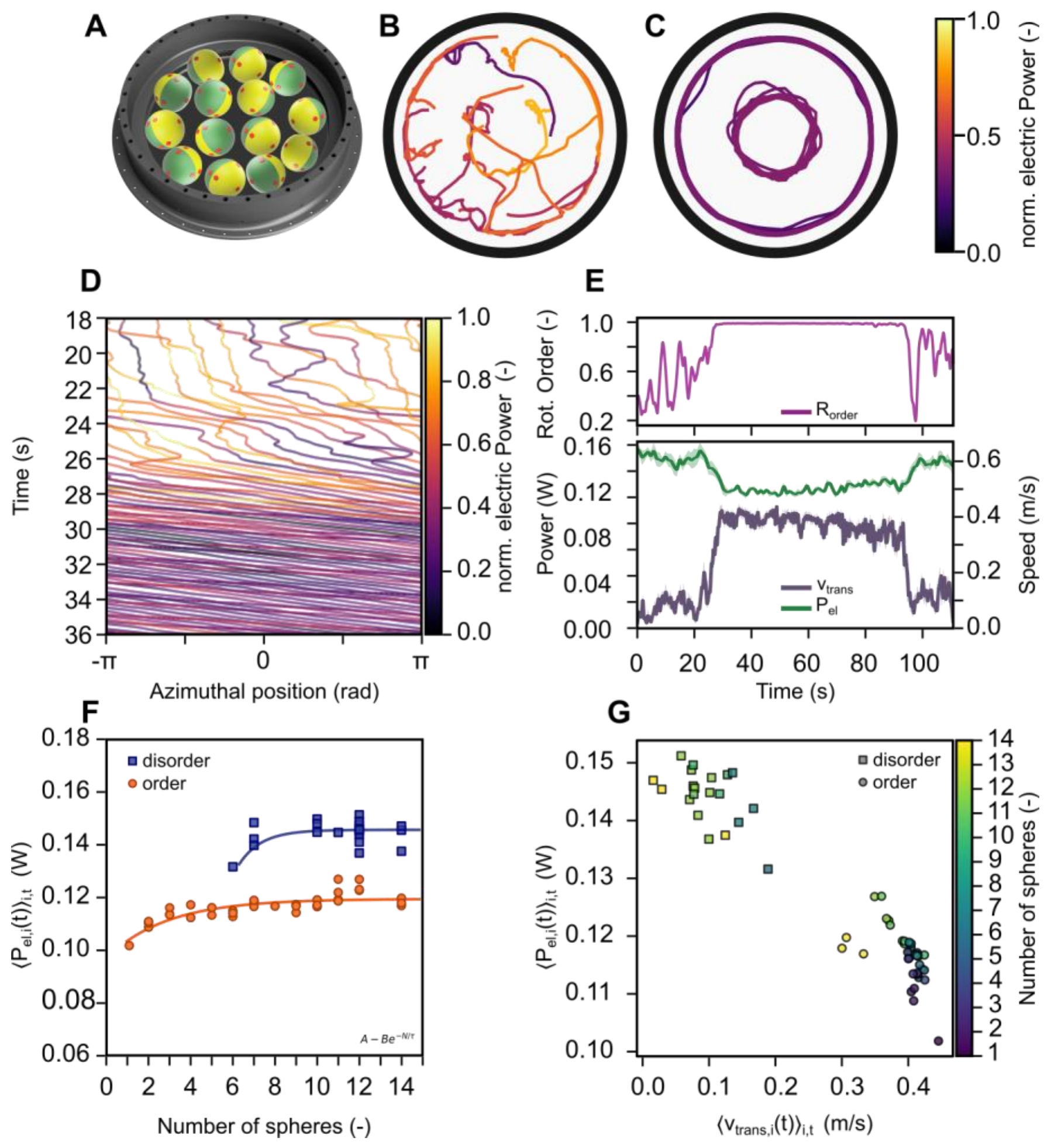

**Fig. 1. The transition to ordered motion reduces energy consumption and increases translational velocity. (A)** Schematic of the active-sphere system in a circular confinement ($R_{conf} = 180\ mm$). Representative $6s$ trajectories in the **(B)** disordered state and **(C)** ordered state ($N_{spheres} = 14$). **(D)** Space-time representation showing the azimuthal position of the active spheres over time. The transition from irregular trajectories to bands with constant slope marks the transition from disordered to ordered motion ($N_{spheres} = 14$). **(B, C, D)** Colors show normalized electrical input power which decreases during transition. **(E)** Time series from a representative experiment ($N_{spheres} = 12$) showing (Top) rotational order, (Bottom) mean translational speed, and mean electrical input power during the transition from disordered to ordered motion. **(F)** Across the investigated population sizes, the mean electrical input power per sphere is consistently higher in the disordered state than in the ordered state. **(G)** Ordered states cluster in a regime of higher translational speed and lower electrical input power; in contrast, disordered states occupy the opposite regimes.

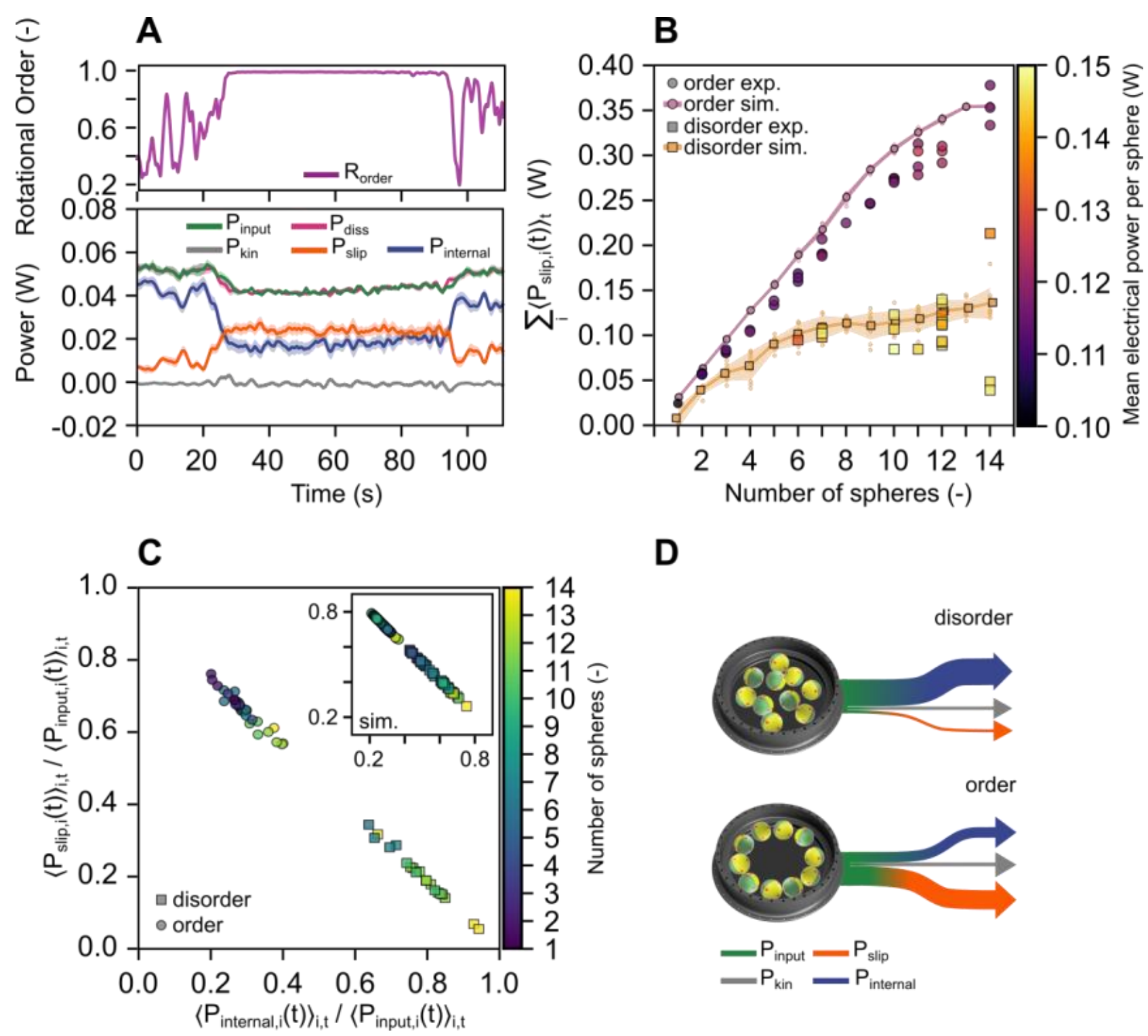

**Fig. 2. Collective ordering reorganizes dissipation from internal to external pathways. (A)** Representative power-budget time series ($N_{spheres} = 12$) during the transition from disordered to ordered motion, showing (Top) rotational order and (Bottom) input power ($P_{input}$), total dissipation ($P_{diss}$), kinetic power ($P_{kin}$), slip dissipation ($P_{slip}$), and internal dissipation ($P_{internal}$). **(B)** Total slip dissipation increases linearly with increasing number of active spheres in the ordered state both in experiments and simulations. **(C)** Ordered (circles) and disordered (squares) states separate into distinct dissipation regimes, with the ordered state characterized by greater slip dissipation and reduced internal dissipation. (Inset) Simulation results show the same separation of dissipation pathways. **(D)** Schematic summary of the inferred power redistribution: in the disordered state, a larger fraction of input power is dissipated internally, whereas collective order shifts dissipation toward external slip at the sphere–substrate interface.

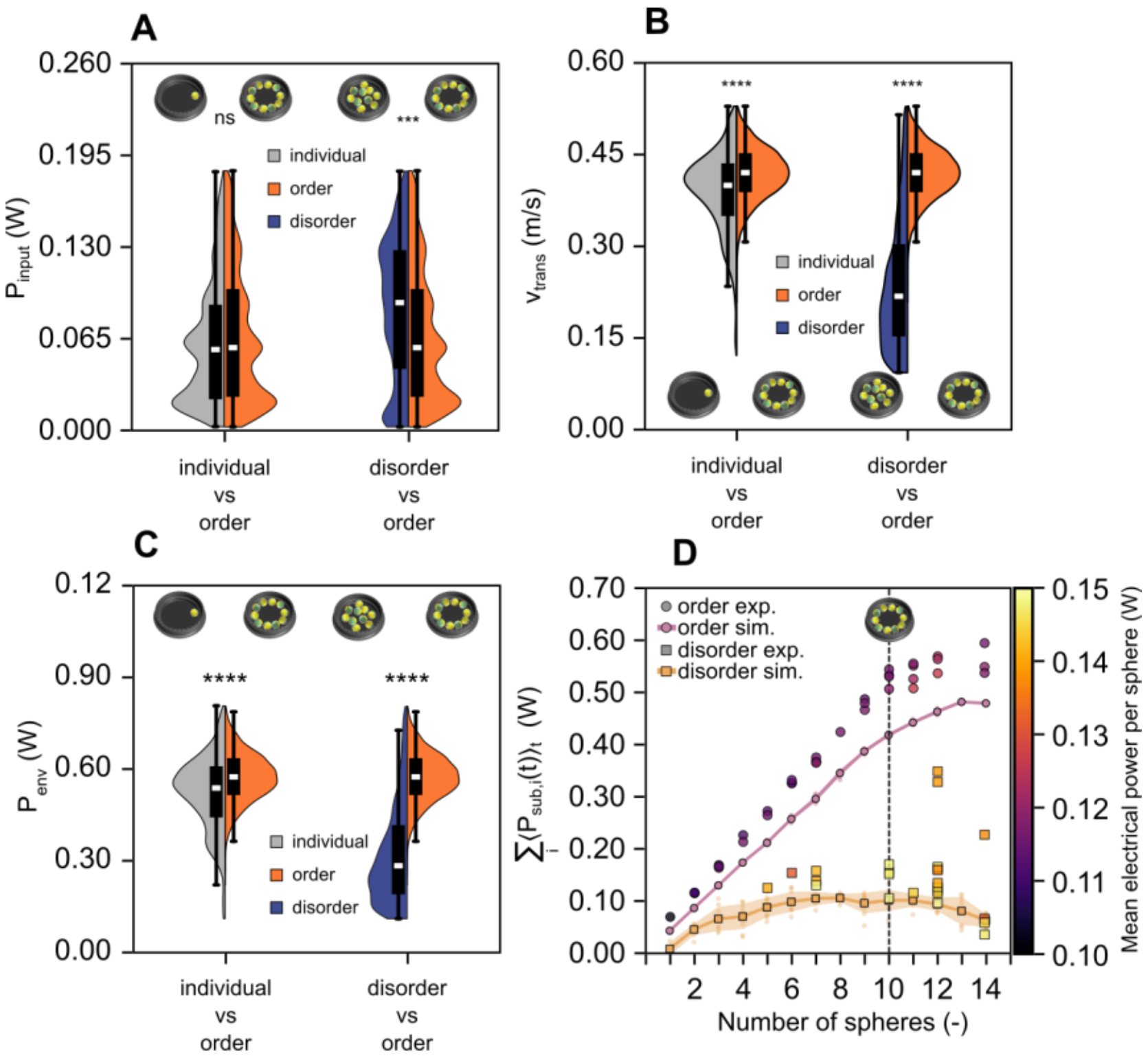

**Fig. 3. Ordered collectives move faster and increase the mechanically transferable power to the environment without increasing input power. (A)** Distributions of input power for isolated spheres and for spheres in disordered and ordered collective states. (**B**) Distributions of translational speed for the same three conditions show a significant increase of the collective compared to the isolated spheres. (**C**) The ordered collective has a significantly higher transferable power to the environment compared to isolated or disordered collectives. **(D)** The total transferable power increases with the number of spheres. Colors indicate the mean electrical input power per sphere. An optimum between electrical input power and transferable power is reached when the outer ring is filled. Simulations reproduce this trend.

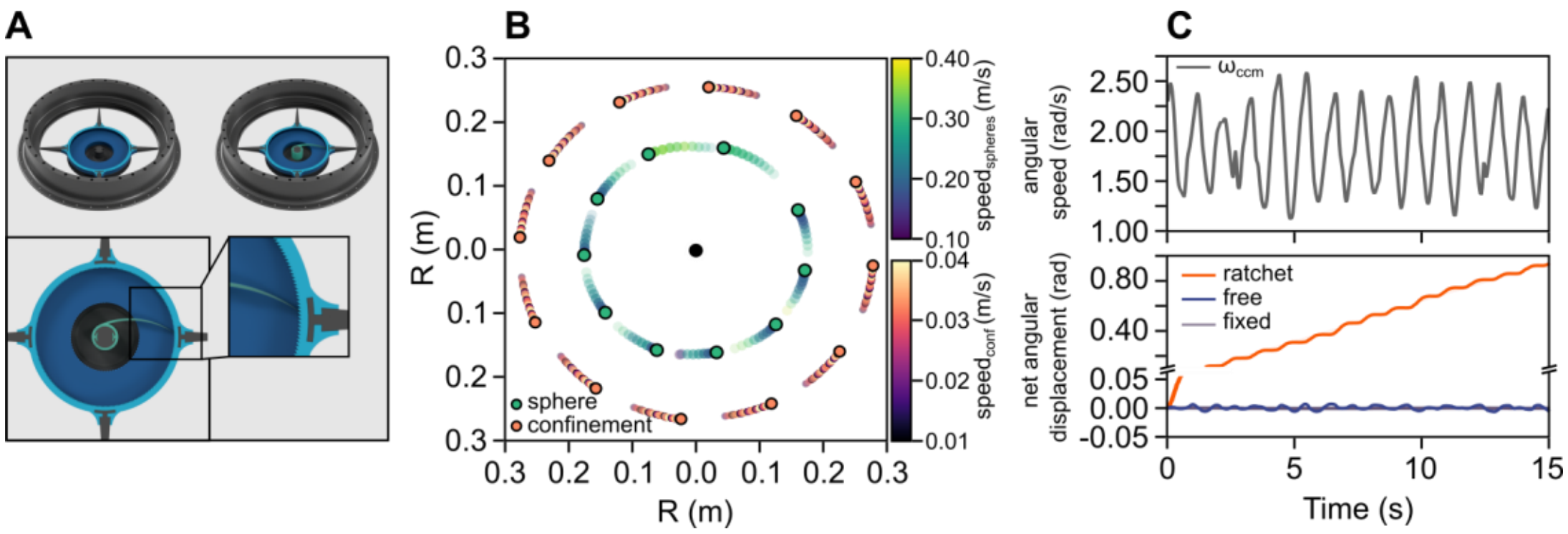

**Fig. 4. A mechanical ratchet converts collective forcing into directional rotation of the confinement. (A)** Schematic of the ratchet mechanism. (Top left) Confinement without pawl. (Top right) Confinement with pawl. (Bottom) Top view showing the ratchet and the interaction between the tooth geometry and the pawl. **(B)** Sphere and confinement marker trajectories. **(C)** (Top) Angular velocity of the collective center of mass for the ensemble of the active spheres. (Bottom) Net angular displacement of a fixed confinement, the oscillating motion of a freely rotating confinement on the ratchet mechanism without a pawl, and the sustained rotation of the confinement on the ratchet mechanism with the pawl engaged. The angular velocity shown in the top panel corresponds to the ratchet case of the bottom panel.

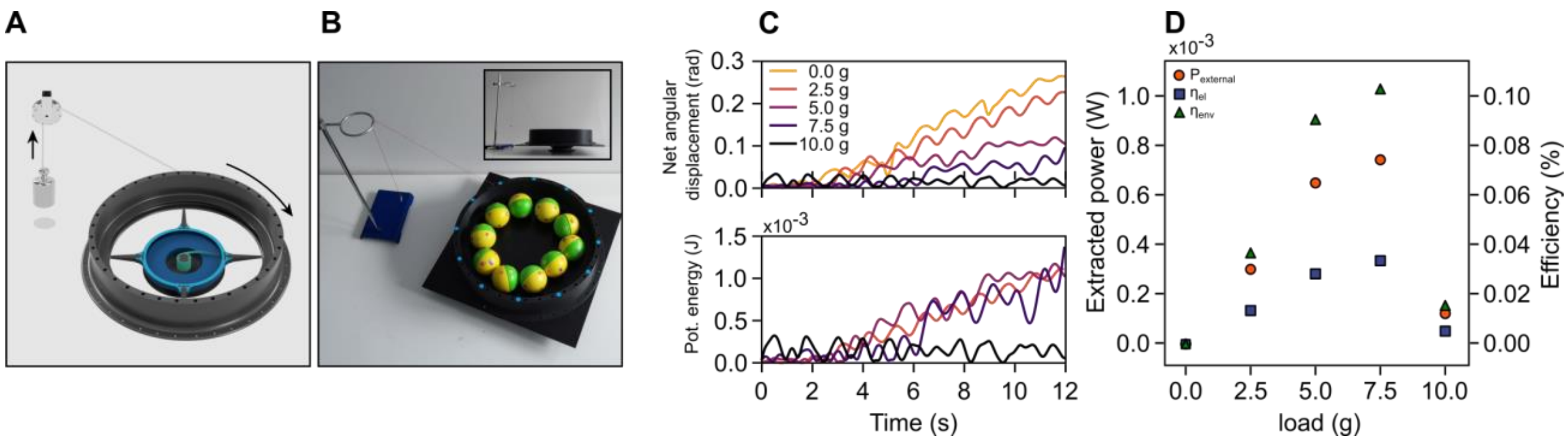

**Fig. 5. The ordered collective performs mechanical work lifting an external load. (A)** Schematic of the active engine, in which rotation of the confinement is coupled to a pulley that lifts a suspended mass. (B) Photograph of the experimental realization. (C) (Top) Net angular displacement of the confinement over time for different applied loads. (Bottom) The corresponding increase in gravitational potential energy of the lifted mass. Increasing the external load reduces the displacement per forcing cycle, but the corresponding gain in potential energy stays constant until the rotation stalls. (D) Extracted mechanical power and efficiencies as a function of the external load.

# Methods

## Experimental system: active spheres

Our active spheres are constructed using as a base body the commercially available motorized toy sold as “weasel ball”, or “beaver ball” or similar by D.Y. Toy **(Supplementary Fig. 1)**. The original spherical shell and internal unbalanced motor assembly are retained, while the additional components are integrated to enable remote operation and measurements.

Each sphere comprises a rigid hollow plastic shell (outer diameter 82 mm) enclosing a single unbalanced electric motor (FA-130RA-2270) that acts, together with additional weight, as an eccentric internal mass. The motor rotates freely at a frequency of approximately 2-3 Hz at 1.5 V around a shaft that spans the two poles of the shell **(Supplementary Fig. 1, Supplementary Fig. 2)**. The efficiency of this small electrical motor peaks at roughly $\eta_{motor,max} = 42\%$ at load torque $C_e$ of 0.0005 Nm. When the assembled sphere is placed on a sufficiently rough and even substrate (here polyethylene, coefficient of friction against the shell $\mu = 0.095$), frictional coupling between the shell and the surface converts internal motor rotation into rolling motion of the sphere, resulting in self-propelled movement [29,30,42].

The total mass of each modified sphere is 138 g, comprising the shell (49 g), motor and mechanical fixations (36 g), additional balancing weights (25 g), battery (18 g), and electronics (10 g). All active spheres are assembled to identical specifications.

Propulsion arises solely from internal actuation, without external forcing or imposed directional or chiral bias, which results in chaotic trajectories when moving freely in isolation solely from their dynamics of the active spheres [29,30,42].

**Electrical power measurement and control**

Powering the motor and measuring its electrical power consumption are handled locally within each active sphere using an integrated printed circuit board (PCB). The PCB carries a microcontroller (XIAO ESP32, Seeed Studio) with Wi-Fi capability, enabling wireless transmission of data to a central Raspberry Pi **(Supplementary Fig. 2)**.

Pulse-width modulation (PWM) is used to compensate for the higher supply voltage of the microcontroller (3.7 V Li-ion battery) such that the effective motor speed matches that of the unmodified system. Throughout all experiments, the PWM duty cycle is fixed and identical across all active spheres; no active feedback or speed control is applied. Batteries are recharged or replaced to full capacity prior to each experiment.

Electrical power supplied to the motor is measured using a current–voltage sensor (INA219, Texas Instruments) mounted on the PCB **(Supplementary Fig. 2 C)**. The sensor measures the motor supply bus voltage and the voltage drop across a calibrated shunt resistor in series with the motor, from which motor current is obtained, together these yield the electrical power, which is computed internally. Only the electrical power delivered to the motor is included in $P_{el}$; power consumption of the electronics itself is excluded.

Power data are acquired by reading the INA219 power register at 150 Hz and internally averaged in non-overlapping blocks to reduce noise and to limit data throughput. Depending on the experiment, block lengths of 5 or 10 samples are used, yielding effective power time series at 30 Hz or 15 Hz, respectively. To reduce Wi-Fi traffic even further, measurement data are collected in 2-s packages and, only then, transmitted wirelessly to the Raspberry Pi for logging.

### Experimental setup and confinement

Experiments are performed in a circular confinement with an inner diameter of 360 mm **(Fig. 1)**. The confinement is fabricated by 3D printing in polylactic acid (PLA) and forms a rigid boundary which is fixed to a 5-mm-thick polyethylene (PE) substrate. At full occupancy, the confinement accommodates a total of 14 spheres, 10 spheres along the outer ring and up to 4 spheres in the central region, with a maximum packing ratio of 0.74.

### Motion tracking

The entire setup is imaged from above at 30 Hz under homogeneous illumination. The image sequences are processed in Fiji[43]. Processing the spheres and reference markers on the shells independently, we use color thresholding to isolate either the spheres, the reference markers on the shells, or, when relevant, reference markers on the confinement **(Supplementary Fig. 3)**. Using the Trackmate[43,44] plugin we identified and tracked the relevant features, yielding time-resolved two-dimensional trajectories for shells and markers.

### Kinematic analysis

Combining the 2D shell trajectories with their known sphere radius, the sphere centers are treated as 3D trajectories with constant height, $x_i(t) = \left(x_i(t), y_i(t), z = R_{sphere}\right)$, where z equals the sphere radius $R_{sphere}$, and $i$ denotes the $i$-th sphere. The translational speed of sphere $i$ is obtained from finite differences of measured trajectories, resulting in the in-plane center-of-mass speed $|v_i(t)|$. In conditions where the confinement is allowed to rotate, the kinetic quantities are evaluated in the co-rotating reference frame with respect to the confinement.

From the 2D trajectories we additionally compute the polar coordinates with respect to the center of the confinement. The collective angular-alignment order parameter is then $R = \left|\frac{1}{N}\sum_i e^{i\theta_i(t)}\right|$, with $R = 1$ for perfectly aligned motion and $R = 0$ for uncorrelated directions. The angular direction of each sphere $\theta_i(t)$ at time t is defined as the angle between its velocity vector and the radial direction originating at the confinement center.

Reference markers assigned to a sphere by a threshold on the radial distance corresponding to $R_{sphere}$ from its center are expressed relative to the sphere center and projected onto the shell's surface. The markers lie predominantly in the proximity of the equator of the dominant rotation axis, where the tangential surface displacement is maximal. Their positions are converted to spherical coordinates and differentiated in time to obtain the angular velocity components. These angular velocity components are combined into a per-marker angular speed on the spherical surface, which is then averaged across all reference markers present at time t to yield a uniform per-sphere angular rotational rate $\omega_i(t)$. This rotational rate is converted into a tangential surface speed by multiplication with the known sphere radius, $v_{rot,i}(t) = R_{sphere} \cdot \omega_i(t)$. The difference between this tangential surface speed and the translational speed of the sphere defines the slip velocity, $v_{\text{slip},i}(t) = v_{\text{rot},i}(t) - |v_i(t)|$.

**Power budget and dissipation decomposition**

The electrical power $P_{el,i}(t)$ supplied to the motor of sphere $i$, is obtained directly from the onboard current–voltage measurements. To convert electrical power to mechanical power $P_{in,i}(t)$, here we use a motor efficiency $\eta_{motor} = 35\%$. To relate this mechanical input power with the  motion, we decompose the energy flow of each sphere into translational, rotational, and dissipative channels using the kinematic quantities defined above.

The translational kinetic energy associated with the center-of-mass motion is

$$E_{trans,i}(t) = \frac{1}{2}\, m\, |v_i(t)|^2,$$

where $v_i(t)$ is the translational velocity of an individual sphere. Its time derivative gives the translational kinetic power

$$P_{trans,i}(t) = \frac{dE_{trans,i}(t)}{dt}.$$

The rotational motion of the shell, characterized by the angular rate $\omega_i(t)$ extracted from surface-marker tracking, carries the rotational kinetic energy

$$E_{rot,i}(t) = \frac{1}{2}\, I\, \omega_i(t)^2,$$

where I is the moment of inertia of a sphere. The associated rotational power is

$$P_{rot,i}(t) = I\, \omega_i(t)\, \frac{d\omega_i(t)}{dt}.$$

The kinetic power corresponding to the motion is therefore

$$P_{kin,i}(t) = P_{trans,i}(t) + P_{rot,i}(t).$$

The relative sliding between the sphere and substrate, quantified by the slip velocity $v_{slip,i}(t)$, produces a frictional power dissipation

$$P_{slip,i}(t) = \mu F_N v_{slip,i}(t).$$

where we experimentally estimate $F_N$ to be $m\,g$. All remaining power not converted into kinetic energy or slip is grouped into an internal dissipation channel, defined by

$$P_{diss,i}(t) = P_{in,i}(t) - P_{kin,i}(t) - P_{slip,i}(t).$$

Not all dissipation corresponds to losses, some can perform mechanically transferable work. Only dissipation mediated by frictional contact with the substrate or the confining boundary can, in principle, be converted into useful mechanical output. We therefore estimate the rate at which mechanical energy can be transferred from each sphere to its environment. In the case of a fixed substrate, this becomes a theoretical upper bound. The maximum power that can be transmitted to the substrate using $F_N \approx m\,g$ is

$$P_{sub,i}(t) = \mu\, m\, g\, |v_{rel,i}(t)|,$$

where $v_{rel,i}(t)$ is the translational speed of sphere i relative to the substrate.

The maximum power that can be transferred to the confining wall through tangential contact during circular motion is approximated as

$$P_{wall,i}(t) = \mu\, m\, \frac{|v_{\theta,i}(t)|^2 * v_{rel,\theta,i}(t)}{(R_{conf} - R_{sphere})},$$

where $v_{\theta,i}$ is the absolute azimuthal velocity of sphere $i$ with respect to the confinement center, $v_{rel,\theta,i}$ is the relative azimuthal velocity with respect to the confinement ($v_{rel,\theta,i} = v_{\theta,i}$, if the confinement is fixed) and $R_{conf}$ denotes the confinement radius. The wall contribution is only considered when a sphere is in direct contact with the boundary, which is identified from its radial position $r_i(t) = R_{conf} - R_{sphere}$; otherwise $P_{wall,i}(t)$ is set to zero.
The sum of these two powers defines an upper bound on the mechanically transferable environmental power,

$$P_{env,i}(t) \leq P_{sub,i}(t) + P_{wall,i}(t)\ .$$

This quantity represents the maximum instantaneous mechanical power that could, in principle, be delivered from the active spheres through substrate and wall interactions.

**Active engine and mechanical load**

To construct a proof of principle active engine powered by collective motion, we convert the circular confinement into a rotary output stage. The confinement is mounted on an axial thrust ball bearing (type 51120; 100 mm inner diameter, 135 mm outer diameter, 25 mm width) to reduce friction and to allow smooth rotation around the vertical axis.

A gear ring fixed to the rotating confinement engages with a removable pawl. With the pawl engaged, the gear–pawl pair acts as a ratchet that blocks one sense of rotation while permitting the other, thereby rectifying the oscillating torque generated by the spheres into intermittent, step-like forward rotation. With the pawl removed, the confinement is free to rotate in both directions and serves as a low-friction reference condition.

The confinement rotation is tracked using the reference markers on the confinement as described earlier. From the positional change of these markers with respect to the confinement center we calculate the angular velocity and acceleration of the confinement, $\omega_{conf}(t)$, $\alpha_{conf}(t)$.

Applying a set of known constant torques $C_{conf}$ to the empty confinement, we determined its moment of inertia from the measured angular acceleration,

$$I_{conf} = \frac{C_{conf}}{\alpha_{conf}}.$$

This calibration allows us to quantify the kinetic energy stored in the rotating confinement as

$$W_{conf}(t) = \frac{1}{2} I_{conf} \left[\omega_{conf}(t)\right]^2,$$

and the corresponding instantaneous mechanical power as

$$P_{conf}(t) = I_{conf}\, \omega_{conf}(t)\, \alpha_{conf}(t).$$

The ratchet converts the oscillatory forcing of the spheres on the confinement into a net directional rotation. Because motion is transmitted through a gear–pawl interface, forward progress occurs in discrete mechanical increments set by the gear pitch: the confinement must advance by at least the angular spacing between two gear teeth, $\Delta\theta_{gear}$, for the pawl to click and allow one step of rotation. Each ratchet step therefore corresponds to a linear lifting distance $h = R_{ratchet}\, \Delta\theta_{gear}$, where $R_{ratchet}$ is the effective radius of the gear. Using $R_{ratchet} = 0.24m$ and $\Delta\theta_{gear} = 1.5°$, the corresponding linear increment to overcome one tooth is approximately 6.3 mm.

To characterize the mechanical output under load, we attached balancing weights of 2.5 g, 5.0 g, 7.5 g, and 10.0 g to the rotary output stage of the ratchet. The weights are lifted via a pulley mechanism that is connected via a string at an attachment radius $R_{attachment} = 0.20\, m$. From the tracked confinement angle $\theta_{conf}(t)$, the lifted distance follows as $s(t) = R_{attachment} \left|\theta_{conf}(t) - \theta_{conf}(t_0)\right|$, and the corresponding gravitational potential energy is $E_{pot}(t) = m_{load}\, g\, s(t)$.

The instantaneous extracted mechanical power is obtained as

$$P_{external}(t) = \frac{dE_{pot}(t)}{dt},$$

and its temporal average $\langle P_{external} \rangle$ quantifies the average extracted power delivered against the external load. We compare this extracted power to two reference power scales: the total electrical input power supplied to the motors, $\langle P_{el} \rangle$, and the mechanically transferable power to the confinement, $\langle P_{env} \rangle$.

This defines two efficiencies

$$\eta_{el} = \frac{\langle P_{external} \rangle}{\sum_i \langle P_{el,i} \rangle}$$

and

$$\eta_{env} = \frac{\langle P_{external} \rangle}{\langle P_{env} \rangle},$$

which respectively quantify the electrical-to-mechanical and mechanical-to-mechanical conversion efficiencies of the collective active engine under load.

## Simulations: active spheres

### Coordinate systems and transformations of reference frames

Three reference frames are used to simulate the dynamics of the active sphere system:

- The Galilean reference frame linked to the substrate on which the sphere rolls and slides
- The reference frame moving with the sphere's center
- The reference frame rotating with the surface of the sphere

The corresponding coordinate systems will be used:

- $(O, \hat{e}_x, \hat{e}_y, \hat{e}_z)$ Cartesian coordinate system with origin on the substrate
- $(O_s, \hat{e}_x, \hat{e}_y, \hat{e}_z)$ Cartesian coordinate system with origin at the sphere's center
- $(O_s, \hat{e}_{x_s}, \hat{e}_{y_s}, \hat{e}_{z_s})$ Cartesian coordinate system rotating with the sphere
- $(O_s, \hat{e}_{r_s}, \hat{e}_{\theta_s}, \hat{e}_{z_s})$ Cylindrical coordinate system with the axis on the motor’s shaft $\hat{e}_{z_s}$
- $(O_s, \hat{r}_e, \hat{e}_{\theta_e}, \hat{e}_{\varphi_e})$ Spherical coordinate system centered around $O_s$

The position of the motor relative to the sphere’s center is

$$\vec{r_e} = R_e \sin\theta_e \cos\varphi_e\, \hat{e}_{x_s} + R_e \sin\theta_e \sin\varphi_e\, \hat{e}_{y_s} + R_e \cos\theta_e\, \hat{e}_{z_s}$$
$$\vec{r_e} = r_{ex_s}\hat{e}_{x_s} + r_{ey_s}\hat{e}_{y_s} + z_e\hat{e}_{z_s}$$
$$\vec{r_e} = r_{ex}\hat{e}_x + r_{ey}\hat{e}_y + r_{ez}\hat{e}_z$$
$$\vec{r_e} = r_{exy_s}\hat{e}_{r_s} + z_e\hat{e}_{z_s}$$

where $R_e$ is the distance of the engine from the sphere’s center, $\theta_e$ and $\varphi_e$ are the angular coordinates of the engine’s position in the spherical coordinate system defined above. The active sphere has an instantaneous rotational rate vector $\vec{\Omega} = \begin{pmatrix} \dot{\varphi} \\ \dot{\theta} \\ \dot{\psi} \end{pmatrix}$, and by using the Euler-Rodrigues formula, the instantaneous rotation matrix, given an infinitesimal time step dt, is

$$R_\alpha = \begin{pmatrix} u^2(1-c)+c & uv(1-c)-ws & uw(1-c)+vs \\ uv(1-c)+ws & v^2(1-c)+c & vw(1-c)-us \\ uw(1-c)-vs & vw(1-c)+us & w^2(1-c)+c \end{pmatrix}$$

where $c = cos\,\alpha$ and $s = sin\,\alpha$, and $\alpha = \vec{\Omega}dt$.

**Dynamics of a single active sphere**

Considering a single active sphere moving on an infinite horizontal plane such that there is no collision, the sphere is forced in constant contact with the plane. The sphere is modeled as a hollow shell that bears an axis $\hat{e}_{z_s}$ on which is placed an arm connected to a motor. The arm is free to rotate around the axis and the other rotations are blocked. Only the motor and the sphere's shell have a mass, $m_e$ and $m_s$, respectively. The mass of the shell is supposed uniform and its thickness is assumed to be negligible.

The motor is located in a non-Galilean reference frame linked to the surface of the sphere via the massless arm to the sphere’s axis (shaft). Then, inertial forces must be taken into account such that

$$\vec{F_e} = -m_e\dot{\vec{\Omega}} \wedge \vec{r_e} - m_e\left[\vec{a_s} + \vec{\Omega} \wedge \left(\vec{\Omega} \wedge \vec{r_e}\right) + 2\vec{\Omega} \wedge (\vec{\omega_e} \wedge \vec{r_e})\right] = \overrightarrow{F_{e,acc}} + \overrightarrow{F_{e,vel}}$$

The force due to the weight of the motor and the shell are respectively

$$\vec{W_e} = m_e\vec{g} \text{ and } \vec{W_s} = m_s\vec{g}$$

Since the sphere is always in contact with the plane, a normal force must be exerted by the plane on the sphere

$$\overrightarrow{F_n} = -\min\left(\left(\overrightarrow{W_e} + \overrightarrow{W_s} + \overrightarrow{F_e}\right) \cdot \hat{e}_z, 0\right) \hat{e}_z.$$

To ensure the displacement of the sphere, we introduce a friction force, modeled as dry friction on the plane

$$\overrightarrow{F_t} = -\mu|\overrightarrow{F_n}| \frac{\vec{v}_{slip}}{|\vec{v}_{slip}|},$$

where $\vec{v}_{slip} = \overrightarrow{u_s} - \vec{\Omega} \wedge R_{sphere}\hat{e}_z$ is the slip velocity and μ is the friction coefficient between the shell and the substrate. The acceleration of the center of the sphere $\overrightarrow{a_s}$ is then obtained by $(m_e + m_s)\,\overrightarrow{a_s} = \overrightarrow{F_t}$

The arm which is holding the motor is only able to rotate around the axis of the sphere. Then, a reaction force must compensate for the other forces on blocked directions

$$\begin{cases} \vec{T} \cdot \hat{e}_{r_s} = -\left(\overrightarrow{F_e} + \overrightarrow{W_e}\right) \cdot \hat{e}_{r_s} - m_e \omega_e^2 \left(\overrightarrow{r_e} \cdot \hat{e}_{r_s}\right) \\ \vec{T} \cdot \hat{e}_{z_s} = -\left(\overrightarrow{F_e} + \overrightarrow{W_e}\right) \cdot \hat{e}_{z_s} \end{cases}$$

On the free-rotating direction, a friction force is opposed to the rotation of the arm around the axis of the sphere

$$\overrightarrow{F_g} = -\mu_g |\vec{T} \cdot \hat{e}_{r_s}| \frac{\overrightarrow{\omega_e}}{|\overrightarrow{\omega_e}|}$$

where $\mu_g$ is the friction factor at the shaft. The torque resulting from these forces around the center of the sphere $O_s$ and around the $\hat{e}_{z_s}$ axis yields

$$\overrightarrow{M_{O_s}}\left(\overrightarrow{F_e}\right) = \overrightarrow{r_e} \wedge \overrightarrow{F_e} = \overrightarrow{r_e} \wedge \overrightarrow{F_{e,acc}} + \overrightarrow{r_e} \wedge \overrightarrow{F_{e,vel}}$$
$$\overrightarrow{M_{O_s}}\left(\overrightarrow{P_e}\right) = \overrightarrow{r_e} \wedge \overrightarrow{W_e} \text{ and } \overrightarrow{M_{O_s}}\left(\overrightarrow{P_s}\right) = \vec{0}$$
$$\overrightarrow{M_{O_s}}\left(\overrightarrow{F_n}\right) = \vec{0}$$
$$\overrightarrow{M_{O_s}}\left(\overrightarrow{F_t}\right) = -R_s \hat{e}_z \wedge \overrightarrow{F_t}$$
$$\overrightarrow{M_{O_s}}\left(\vec{T}\right) = \overrightarrow{r_e} \wedge \vec{T} = \overrightarrow{r_e} \wedge \overrightarrow{T_{acc}} + \overrightarrow{r_e} \wedge \overrightarrow{T_{vel}}$$
$$\mathrm{M}_{z_s}\left(\overrightarrow{F_g}\right) = -\mu_g \left|\vec{T} \cdot \hat{e}_{r_s}\right| sign(\omega_e)$$

A small deformation of the sphere on the ground is accounted for, resulting in a resistive torque applied on the contact surface between the sphere and the floor such that

$$\overrightarrow{M_{floor}} = -\frac{2}{3}\mu\left(\frac{2F_n^4}{\pi E_s}\right) sign(\dot{\psi})\hat{e}_z,$$

where $E_s$ is the Young modulus of the shell. Upon several tests, we stress that such a torque has minimal impact on the sphere's dynamics.

The axial vector and the azimuthal vector are computed to re-write the inertial forces on the shaft

$$\hat{e}_{r_s} = \frac{\overrightarrow{r_e} - \overrightarrow{r_e} \cdot \hat{e}_{z_s}}{|\overrightarrow{r_e} - \overrightarrow{r_e} \cdot \hat{e}_{z_s}|} \text{ and } \hat{e}_{\theta_s} = \hat{e}_{z_s} \wedge \hat{e}_{r_s} = \begin{pmatrix} t_x \\ t_y \\ t_z \end{pmatrix}.$$

Apart for a sign, the angular acceleration terms of the inertial force are given by

$$\overrightarrow{T_{acc+\theta}} = m_e \vec{\Omega} \wedge \overrightarrow{r_e} = m_e \begin{pmatrix} \mathrm{r_{ez}} \ddot{\theta} - \mathrm{r_{ey}} \ddot{\psi} \\ -\mathrm{r_{ez}} \ddot{\varphi} + \mathrm{r_{ex}} \ddot{\psi} \\ \mathrm{r_{ey}} \ddot{\varphi} - \mathrm{r_{ex}} \ddot{\theta} \end{pmatrix}.$$

Considering the modeling assumptions, there is no reaction force on $\hat{e}_{\theta_s}$. It is then possible to link the moment of inertia of the sphere and the torques applied to the sphere

$$J_s \dot{\vec{\Omega}} = \overrightarrow{M_{O_s}}(\overrightarrow{F_t}) + \overrightarrow{M_{O_s}}(-\overrightarrow{T_{vel}}) - \overrightarrow{r_e} \wedge \overrightarrow{T_{acc}} - M_{z_s}(\overrightarrow{F_g}) \hat{e}_{z_s},$$

where $J_s$ is the moment of inertia of the shell, $\overrightarrow{T_{vel}}$ is the tension at the shaft due to the centrifugal acceleration and $\overrightarrow{T_{acc}}$ is a consistent projection of $\overrightarrow{T_{acc+\theta}}$. For the rotation of the motor, it yields

$$m_e |\overrightarrow{r_e} \wedge \hat{e}_{z_s}| \dot{\omega}_e = \overrightarrow{M_{O_s}}(\overrightarrow{F_e}) \cdot \hat{e}_{z_s} + \overrightarrow{M_{O_s}}(\overrightarrow{W_e}) \cdot \hat{e}_{z_s} + M_{z_s}(\overrightarrow{F_g}) + \eta_{motor} C_{el},$$

where $\eta_{motor}$ is the efficiency of the motor and $C_{el}$ its electric torque. To account for the mechanical limitations of the real system, we limit the torque transferred to the motor to a maximum $\eta_{motor} C_{el,max}$, enforced if the power transferred to the motor divided by its angular velocity overcomes such a threshold. Hence $C_e = \min\left(\frac{P_{el}}{\omega_{el}}, \eta_{motor} C_{el,max}\right)$. Moreover, the engine is considered locked to the shaft when it reaches a maximal velocity.

**Interaction between two spheres and with the external confinement wall**

The only interaction that are considered here are partially inelastic collisions between two spheres and between a sphere and the confinement wall. The confinement wall is assumed to be fixed to the plane reference frame (equivalent to an infinite inertia). All the spheres have the same masses, radius, and distance between the sphere's center and radius. Variables before and after collisions are denoted by indices b and a, respectively. The normal unit vector $\hat{n}_{cp}$ is going from sphere 1 to sphere 2. The restitution coefficient ε links relative velocity between two spheres before and after a collision such that

$$\overrightarrow{u^a_{s_1,s_2}} \cdot \hat{n}_{cp} = -\varepsilon \overrightarrow{u^b_{s_1,s_2}} \cdot \hat{n}_{cp}.$$

The collision between two spheres or a sphere and the confinement wall imply the exchange of an impulse $j_{s-i}$

$$\overrightarrow{u^a_{s_1}} = \overrightarrow{u^b_{s_1}} + \frac{j_{s-i}}{m_{e_1}} \hat{n}_{cp} \text{ and } \overrightarrow{u^a_{s_2}} = \overrightarrow{u^b_{s_2}} - \frac{j_{s-i}}{m_{e_2}} \hat{n}_{cp}$$

where i depends on the nature of the second object (sphere or wall)

$$j_{s-s} = -\frac{(\varepsilon+1)\overrightarrow{u^b_{s_1,s_2}} \cdot \hat{n}_{cp}}{1/m_{e_1} + 1/m_{e_2}} \text{ and } j_{s-w} = -\frac{(\varepsilon+1)\overrightarrow{u^b_{cp,s}} \cdot \hat{n}_{cp}}{1/m_e}$$

To prevent interpenetrations between the spheres, they are set back by a small distance (1% of their radius) after the collisions. It is assumed that there is no slip on the plane of collision. So, the rotation of the sphere is directly linked to the velocity of its center by

$$R_{sphere}\overrightarrow{\Omega^a} = \hat{e}_z \wedge \overrightarrow{u^a_s}$$

This relation can only be justified for φ and θ. For ψ, an impulsive friction force is set on contact point. This force is given by

$$\vec{F}_{\mu_{shell}} = -\mu_{shell}\left|\left(m_{e_1}\overrightarrow{a_{s_1}} - m_{e_2}\overrightarrow{a_{s_2}}\right) \cdot \hat{n}_{cp}\right| \frac{\overrightarrow{U_1} - \overrightarrow{U_2}}{\left|\overrightarrow{U_1} - \overrightarrow{U_2}\right|},$$

where $\overrightarrow{U_\iota} = \left[\left(\overrightarrow{u_{s_\iota}} + (-1)^{i+1} R_{sphere}\vec{\Omega} \wedge \hat{n}_{cp}\right) \cdot \hat{t}_{cp}\right]\hat{t}_{cp}$ is the contact point velocity relative to the plane, $\hat{t}_{cp}$ is the tangential vector in the collision plan and orthogonal to $\hat{e}_z$ and $\mu_{shell}$ is the friction between interacting shells. The rotation velocity around $\hat{e}_z$ is then modified as

$$J_{z_{1,2}}\dot{\psi}^a_{1,2} = J_{z_{1,2}}\dot{\psi}^b_{1,2} + \left(\int_{t^b}^{t^a} R_{sphere}\hat{n}_{cp} \wedge \vec{F}_{\mu_{shell}} dt\right) \cdot \hat{e}_z$$

where $J_z$ is the moment of inertia of the sphere around $\hat{e}_z$. The friction coefficient $\mu_{shell}$ is replaced by a new coefficient $\varepsilon_\omega$ which models time integration. So

$$J_{z_{1,2}}\dot{\psi}^a_{1,2} = J_{z_{1,2}}\dot{\psi}^b_{1,2} + \varepsilon_\omega\left(R_{sphere}\hat{n}_{cp} \wedge \vec{F}_{\mu_{shell}}\right) \cdot \hat{e}_z$$

Finally, the rotation of the motor is affected by the rotation changes of the sphere, and the transfer of angular momentum on the sphere's engine is modeled as

$$\omega_e^a = \omega_e^b + \varepsilon_\varphi \left(\overrightarrow{\omega_s^a} - \overrightarrow{\omega_s^b}\right) \cdot \hat{e}_{z_s},$$

where $\varepsilon_\varphi \in [-1,1]$

**Energy transfer**

The sphere dissipates energy by moving on the floor. This dissipation is due to the friction force on the floor. Within the dynamical model, we compute it as

$$P_{floor} = \overrightarrow{F_t} \cdot \vec{\text{v}}_{slip}$$

The transfer of energy from the motor to the shaft, hence the surface of the sphere is done by the friction force on the rotation gear and the exchanged power is given by

$$P_{gear} = |\overrightarrow{F_g}|\omega_e|\overrightarrow{r_e}|$$

**Numerical method**

The script is written with Python (version 3.13.0) and some of its additional libraries (NumPy 2.1.3). Here we explain the architecture briefly. Given initial conditions and parameter values, the dynamics of the system are firstly computed without accounting for collision for one time step. At the end of each time step, collisions are taken into account and, if two spheres or a sphere and the confinement wall collide, the trajectories are corrected. For the independent dynamics of the spheres, equations are discretized in time. This leads to a system of second order differential equations per each sphere, which equations are not analytically solvable. To solve the system of differential equations, we isolate second order time derivatives. The function numpy.linalg.solve is then used to isolate angular acceleration terms. Once all the second order time derivatives are isolated, we integrate over time with a Runge-Kutta fourth-order method. Collisions are considered as instantaneous events at the end of each time step.